\documentclass[journal,10pt]{IEEEtran}
\usepackage{graphicx}
\usepackage{diagbox}
\usepackage{multirow}
\usepackage{balance}

\PassOptionsToPackage{hyphens}{url}
\usepackage{hyperref}
\hypersetup{colorlinks, citecolor=green, filecolor=pink, linkcolor=red, urlcolor=blue }

\begin{document}

\title{Blockchain-enabled Internet of Medical Things to Combat COVID-19}

\author{Hong-Ning Dai,~\IEEEmembership{Senior Member,~IEEE,} 
		Muhammad Imran,~\IEEEmembership{Senior Member,~IEEE,}
		Noman Haider,~\IEEEmembership{Member,~IEEE}
		
		\thanks{H.-N. Dai is with Faculty of Information Technology, Macau University of Science and Technology, Macau. e-mail: hndai@ieee.org.}
		
		\thanks{M. Imran is with College of Applied Computer Science, King Saud University, Riyadh, Saudi Arabia. email: dr.m.imran@ieee.org.}
		
		\thanks{N. Haider is with College of Engineering and Science at Victoria University, Australia. email: noman90@ieee.org.}%
}

\markboth{IEEE Interent of Things Magazine}%
{Dai \MakeLowercase{\textit{et al.}}: Bare Demo of IEEEtran.cls for IEEE Journals}

%

\maketitle

\begin{abstract}

We are experiencing an unprecedented healthcare crisis caused by the newly-discovered corona-virus disease (COVID-19). The outbreaks of COVID-19 reveal the frailties of existing healthcare systems. Therefore, the digital transformation of healthcare systems becomes an inevitable trend. During this process, the Internet of Medical Things (IoMT) plays a crucial role while intrinsic vulnerabilities of security and privacy deter the wide adoption of IoMT. In this article, we present a blockchain-enabled IoMT to address the security and privacy concerns of IoMT systems. We also discuss the solutions brought by blockchain-enabled IoMT to COVID-19 from five different perspectives. Moreover, we outline the open challenges and future directions of blockchain-enabled IoMT.

\end{abstract}

\begin{IEEEkeywords}
Blockchain, Internet of Medical Things, COVID-19, Security, Privacy
\end{IEEEkeywords}

\IEEEpeerreviewmaketitle

\section{Introduction}

\IEEEPARstart{W}{e} have experienced an unprecedented health crisis due to the pandemic of the newly-discovered corona-virus disease (COVID-19). The COVID-19 crisis has caused severe impacts on both the socio-economy and healthcare systems. Our normal life is affected by city lockdown and flight cancellation. Moreover, the eruption of the disease is not only carrying intense pressure to healthcare professionals (such as doctors and nurses) but also revealing the vulnerabilities of healthcare systems. Healthcare workers have to work overtime for months to deal with overwhelming infectious patients in hospitals. Senior citizens and chronic patients are required to stay at home to avoid possible infections in healthcare institutions.

Fortunately, the recent advances in Information and communications technology (ICT) such as the Internet of Things (IoT) and artificial intelligence (AI) bring us opportunities to win the battle against the COVID-19 crisis. In particular, the availability of diverse bio-sensors, medical devices, and wearable devices as well as the advances of wireless communications technologies lead to the advent of the internet of medical things (IoMT)~\cite{MShen:MNET19}. IoMT can be widely adopted in healthcare scenarios such as telemedicine, pandemic quarantine, social distancing, and smart hospitals. IoMT has generated massive medical and healthcare data from diverse IoMT devices and medical facilities in healthcare institutions. After analyzing the IoMT data, we can diagnose and identify diseases, consequently giving appropriate treatments. However, IoMT is also faced with security and privacy vulnerabilities, which deter healthcare agencies from widely adopting IoMT. 

The emerging blockchain technologies can potentially address the security and privacy issues in IoMT. Integrated with security mechanisms such as asymmetric cryptographic schemes and digital signature, blockchain can provide IoMT with certain security protection. Moreover, the decentralization of blockchain systems also mitigates the risks of the failures due to single-point-failure and malicious attacks. After introducing privacy-preservation mechanisms, such as homomorphic obfuscations and differential privacy, blockchain can also preserve the privacy of IoMT data. Moreover, blockchain has intrinsic merits, such as traceability and immutability, which can further improve the data provenance of IoMT. Therefore, blockchain is an ideal carrier for IoMT. As a result, the in-depth integration of blockchain and IoMT can further enhance the IoMT systems.

In this article, we investigate the integration of blockchain and IoMT, which is named as blockchain-enabled IoMT. We explore the opportunities brought by blockchain-enabled IoMT, especially in combating COVID-19. In particular, we first briefly review blockchain technologies and IoMT. We then present an architecture of blockchain-enabled IoMT and discuss the opportunities brought by blockchain-enabled IoMT. We next analyze the solutions of blockchain-enabled IoMT to COVID-19 from five perspectives including 1) tracing pandemic origin, 2) quarantine and social distancing, 3) smart hospital, 4) medical data provenance, and 5) remote healthcare and telemedicine. Moreover, we also discuss open challenges and future directions in blockchain-enabled IoMT. 









\section{Overview of Blockchain and IoMT}
This section presents an overview of blockchain and Internet of Medical Things.

\subsection{Blockchain} 

\begin{figure}[t]
\centering
\includegraphics[width=8.6cm]{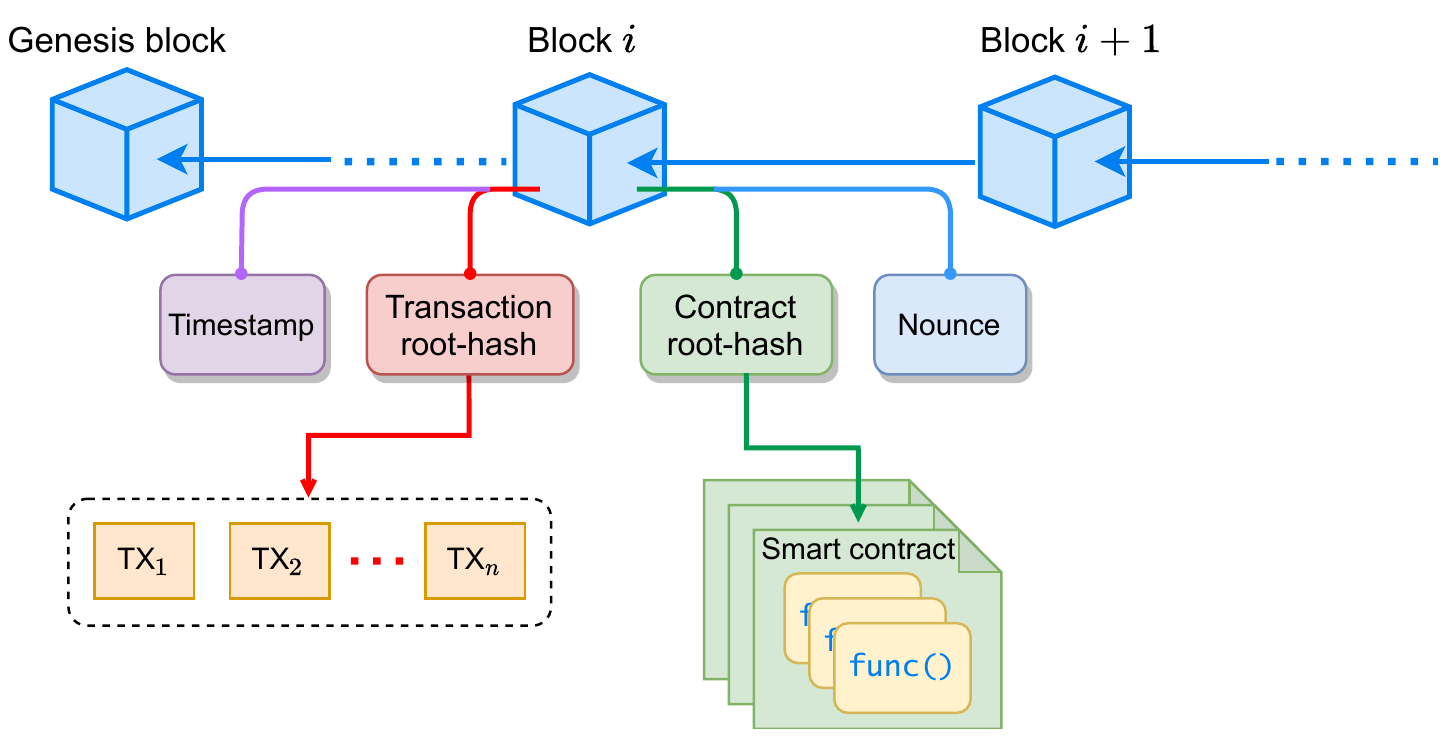}
\caption{Blockchain consisting of a number of blocks.}
\label{fig:blockchain}
\end{figure}

Blockchain consists of a number of blocks as shown in Fig.~\ref{fig:blockchain}. In this structure, except for the first block (i.e., genesis block), every block is essentially connected to its previous block through a backward reference, which is the hash value of its previous block. For instance, block $i+1$ contains the hash value of its previous block $i$. Each block also contains other data fields, such as a timestamp, a cryptographic nonce (a.k.a. a random number used once), a root hash of all the transactions, and a root hash of all the contracts~\cite{dai-iotj-2019}. The fixed-size root hash values of both transactions and contracts are nearly immutable since even one-bit modification can result in a totally different hash value. 

Besides blockchain, smart contracts are another disrupting technology driven by the advent of blockchain. Smart contracts running on top of blockchain can automate the execution of contractual terms and clauses when certain conditions (e.g., delivery delay or breach of contract) are triggered. After being compiled into bytecode (or machine code) and stored in blockchains, smart contracts are immutable due to the immutability of blockchain and the root hash of all the smart contracts. Smart contracts can simplify the administration process, enhance the efficiency of business activities, mitigate the potential risks~\cite{ZZheng:FGCS2020}.

Blockchain technologies have the following key features: (i) \emph{Immutability} means that falsifications on data stored in blockchain are extremely difficult because any slight modifications can lead to invalid data (invalid hash values). (ii) \emph{Non-repudiation} of transactions can be assured by digital signatures, asymmetric cryptographic algorithms, and distributed consensus mechanisms. (iii) \emph{Traceability} implies that the origins of data can be traced through analyzing the publicly available blockchain data with the associated timestamps. (iv) \emph{Decentralization} of blockchain allows the transactions to be validated by the majority of peers distributed in the entire system without the central agency. In this way, the bureaucratic cost can be saved and the system reliability can be enhanced.

Blockchain systems can be typically categorized into three types~\cite{Ahmed:MC20}: 1) public (permissionless) blockchains; 2) private (permissioned) blockchains and 3) consortium blockchains. Public blockchains including Bitcoin, Ethereum, and EOSIO can be publicly accessible by every user in the blockchain network while private blockchains have strict access control to limit the accessibility of users. Consortium blockchains lie between permissionless and permissioned blockchains. Typically, public blockchains have poorer scalability than private blockchains due to the lower throughput of consensus algorithms where the throughput denotes the number of transactions being confirmed every second. Consortium blockchains achieve the performance lower than private blockchains while higher than public blockchains.

\subsection{Internet of Medical Things}
\label{subsec:IoMT}

\begin{figure}[t]
\centering
\includegraphics[width=8.8cm]{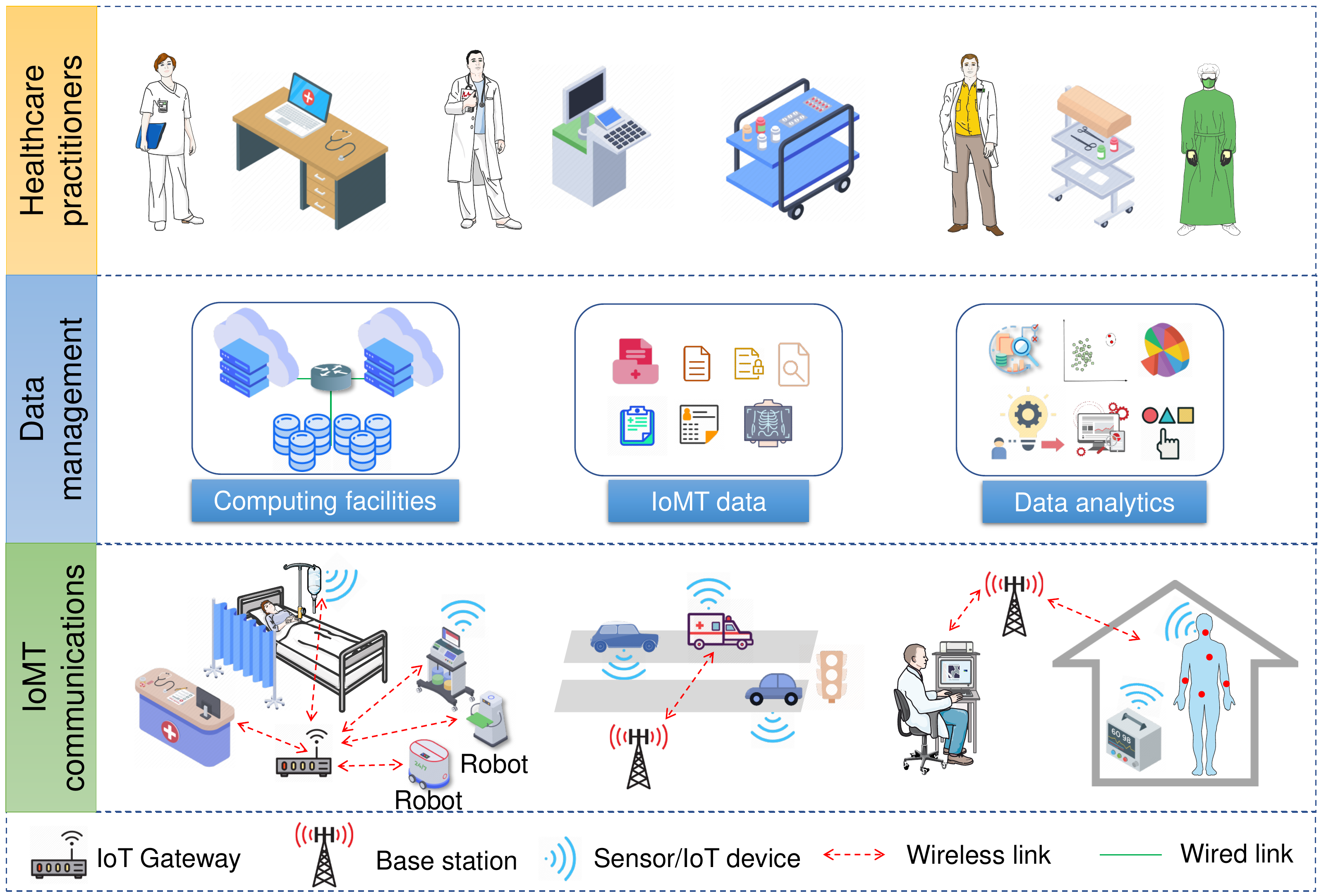}
\caption{Overview of IoMT.}
\label{fig:IoMT}
\end{figure}

The recent advances in bio-sensors, medical devices and communications technologies lead to the advent of IoMT, which can be deployed in diverse healthcare scenarios such as telemedicine, remote rehabilitation, and pandemic quarantine. IoMT that connects diverse medical devices and facilities with healthcare institutions has generated massive heterogeneous medical data. Through analyzing the massive IoMT data, medical professionals (such as doctors and nurses) can diagnose, identify, and make appropriate treatments. Fig.~\ref{fig:IoMT} presents an overview of the existing IoMT systems. In this architecture, diverse IoMT devices generate massive IoMT data, which can be collected, processed, and analyzed, thereby being used by healthcare practitioners~\cite{Qadri:CST20}.

Although IoMT has the potential to offer reliable and effective healthcare services to both patients and medical practitioners, the advent of IoMT also poses the following challenges: 1) \emph{absence of interoperability} across different IoMT sectors; 2) \emph{privacy and security} vulnerabilities of IoMT devices and systems. IoMT consists of diverse biomedical sensors, medical devices, IoT gateways, and base stations, thereby leading to the heterogeneity of IoMT systems. Meanwhile, the diversity of IoMT also exhibits the diversity of various wireless protocols, such as Near-field communication (NFC), Bluetooth Low Energy (BLE), Low-Power Wireless Personal Area Networks (LoWPAN), LoRa and NB-IoT. The heterogeneity of decentralized IoMT systems results in the poor interoperability across different systems, consequently forming a number of information silos. Thus, it is difficult to exchange medical information across different medical facilities and institutions. However, medical information sharing is crucial for medical professionals, especially in prophylaxis and treatment of pandemic outbreaks, such as COVID-19. 

Moreover, IoMT is also faced with rising concerns on security and privacy~\cite{Boudagdigue:TIFS20}. On the one hand, biomedical sensors and medical devices that are often resource-limited (i.e., low computational capability and limited battery power) have intrinsic vulnerabilities to malicious attacks, such as wiretapping, jamming, backdoor, and worm attacks. On the other hand, IoMT data is quite privacy-sensitive compared with other types of IoT data. Particularly, IoMT data often requires to be outsourced to remote cloud servers, which are owned by third parties. During the data collection, processing, and analytics of IoMT data, the data privacy of patients may be intentionally or accidentally leaked.


\section{Blockchain-enabled IoMT}

This section presents an architecture of blockchain-enabled IoMT and discusses the opportunities brought by blockchain-enabled IoMT.

\subsection{Architecture of blockchain-enabled IoMT}

\begin{figure}[t]
\centering
\includegraphics[width=8.8cm]{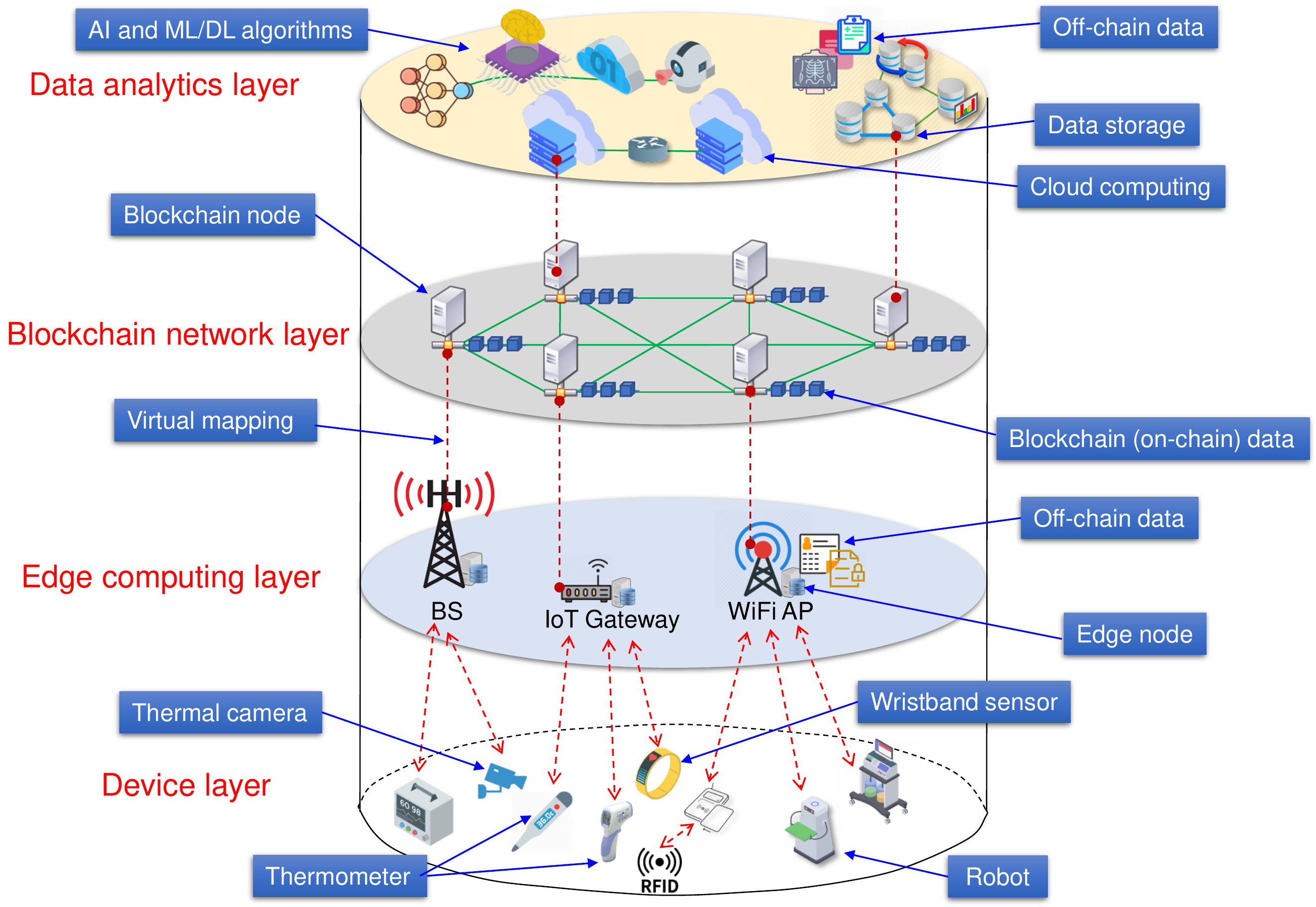}
\caption{Architecture of blockchain-enabled IoMT consisting of device layer, edge computing layer, blockchain network layer and data analytics layer.}
\label{fig:BC-IoMT}
\end{figure}

The integration of blockchain with IoMT can potentially address security and privacy concerns. We name such integration of blockchain with IoMT as blockchain-enabled IoMT. Fig.~\ref{fig:BC-IoMT} presents an architecture of blockchain-enabled IoMT. In this architecture, there are four layers from bottom to up: 1) device layer, 2) edge computing layer, 3) blockchain network layer and 4) data analytics layer. 

At the bottom of this multi-layer framework, there are diverse IoMT devices, such as heat sensors, thermal cameras, wristband sensors, thermometers, and RFID tags in the device layer. In the middle, the edge computing layer is essentially an integration of communication networks and edge computing facilities. Edge computing nodes deployed at base stations, WiFi access points (APs), and IoT gateways can collect and preprocess IoMT data. Moreover, the blockchain network layer plays a crucial role as a middleware to offer trustworthy management of diverse resources across the underlying layers. The data analytics layer on the top consists of cloud computing facilities, data storage servers, and AI and Machine Learning (ML) or Deep Learning (DL) algorithms. It is worth mentioning that either edge computing facilities in the edge computing layer or entities in the data analytics layer are all mapped to nodes in the blockchain network layer. In this manner, blockchain-enabled IoMT can achieve effective authentication and access control on both the edge computing layer and the blockchain network layer.

This multi-layer architecture has the following merits:
\begin{itemize}
    \item \emph{Offering the abstraction} from underlying layers. The edge computing layer and the blockchain network layer essentially play a role as a middleware so that the complexity of IoMT devices and the heterogeneity of IoMT communications can be hidden. Meanwhile, the blockchain network layer can provide other applications (such as data analytics applications) with blockchain-based services so as to facilitate application development.
    \item \emph{Improving the interoperability} of IoMT systems. The blockchain network layer leverages blockchain's built-in overlay peer-to-peer (P2P) network to connect different IoMT sub-networks throughout the entire IoMT system. In this way, the fragmented IoMT components are integrated as a whole so as to offer seamless services to other applications (e.g., data analytics). As a result, the interoperability of IoMT systems can be improved.
\end{itemize}

We next discuss the opportunities brought by blockchain-enabled IoMT.

\subsection{Opportunities of blockchain-enabled IoMT}
As discussed in Section~\ref{subsec:IoMT}, IoMT is faced with security and privacy concerns. Blockchain-enabled IoMT has brought opportunities to address these challenges in IoMT. We next analyze the opportunities brought by blockchain-enabled IoMT in the following aspects.

\subsubsection{Security improvement of IoMT}
The introduction of blockchain to IoMT can significantly improve the security of IoMT. First, the built-in security mechanisms of blockchain, such as asymmetric encryption/decryption schemes as well as the digital signature can offer certain protection to IoMT data. Second, the integration of blockchain with other security schemes, such as authentication and access control can further enhance system security. Third, smart contracts embedded in IoT devices can automatically trigger the auto-upgrading programs to upgrade IoT devices firmware automatically, thereby enhancing the system security~\cite{ZZheng:FGCS2020}. Moreover, the decentralization of blockchain can also mitigate the risks of system failures due to  single-point-failures or other malicious attacks (e.g., distributed denial-of-service attack), thereby improving system security and reliability.



\begin{figure*}[t]
\centering
	\includegraphics[width=16.8cm]{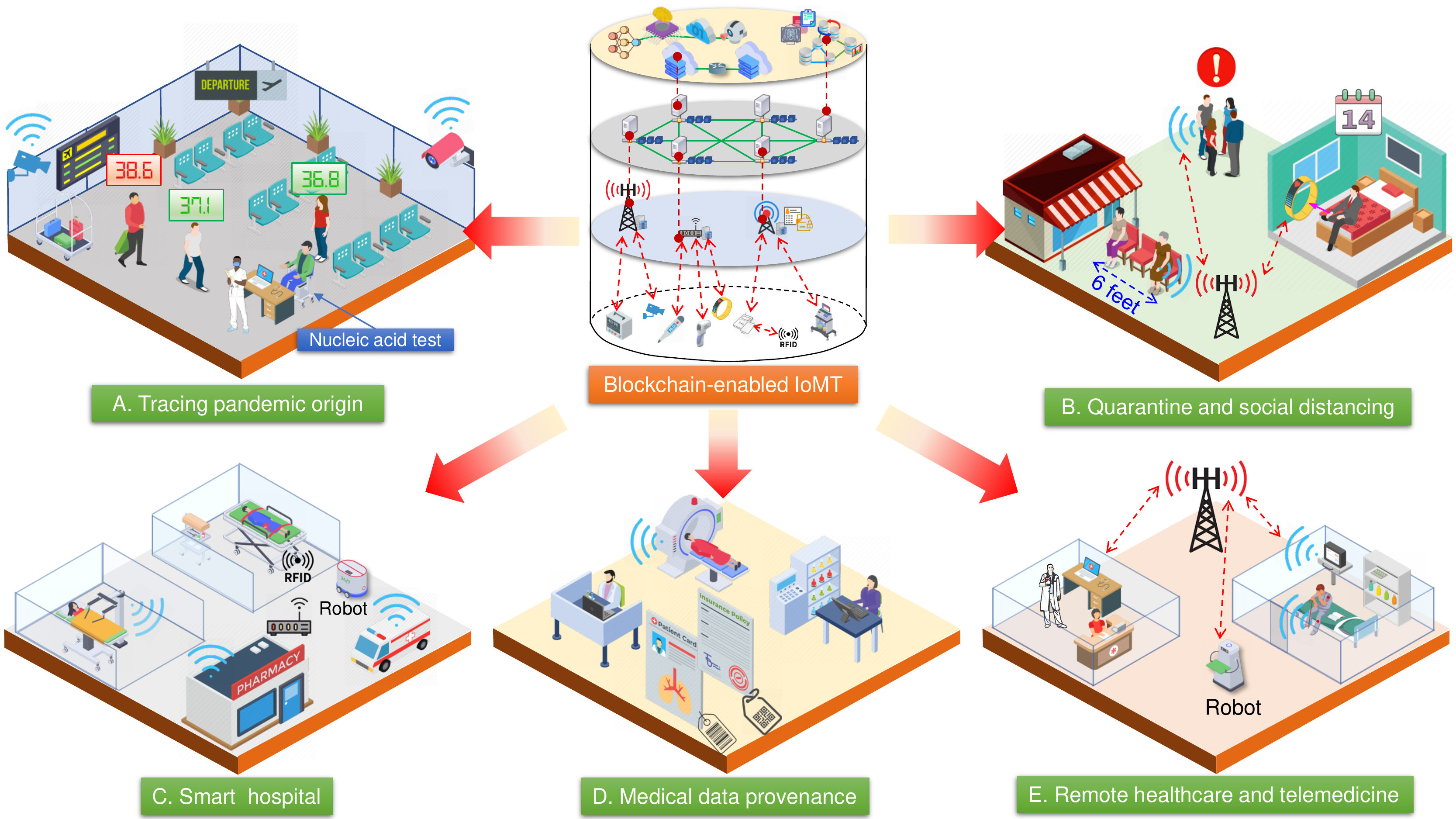}
	\caption{Solutions of blockchain-enabled IoMT to COVID-19.}
	\label{fig:scenario}
\end{figure*}

\subsubsection{Privacy preservation of IoMT data}
Blockchain intrinsically can achieve somewhat privacy preservation through hiding blockchain account addresses and encrypting blockchain transaction data. The integration of blockchain with other privacy-preservation schemes, such as homomorphic obfuscations and other cryptographic schemes, blockchain-enabled IoMT can provide users with better privacy preservation. Moreover, as shown in the architecture of blockchain-enabled IoMT (in Fig.~\ref{fig:BC-IoMT}), edge computing nodes in the edge computing layer can undertake data collection and processing tasks in approximation to users. In this manner, the privacy-sensitive IoMT data can be stored and processed locally before being outsourced to remote clouds. Consequently, data privacy can be preserved via the introduction of the edge computing layer.

\subsubsection{Traceability of IoMT data}
In blockchain-enabled IoMT, IoMT data can be categorized into on-chain data and off-chain data depending on whether the IoMT data is stored at the blockchain or not. The on-chain data stored at blockchain is essentially traceable throughout the entire system. The traceability and non-repudiation of on-chain data can be achieved by the decentralized consensus algorithms and asymmetric cryptographic schemes (i.e., digital signature) of blockchain. However, it is impractical to store all IoMT data into blockchain considering the massive volumes of IoMT data, especially for medical images and videos. Therefore, IoMT data such as images and videos should be stored in an off-chain manner while blockchain may only store meta-data or hash values of the off-chain IoMT data. Moreover, the introduction of digital signatures and access control schemes can improve the traceability of off-chain IoMT data in blockchain-enabled IoMT. For example, storing hash values of off-chain IoMT data in blockchain can also ensure the traceability of IoMT data while saving the storage cost at blockchain.

We next discuss the solutions blockchain-enabled IoMT to address the limitations of existing IoMT systems and combat the COVID-19 crisis.

\section{Solutions of blockchain-enabled IoMT to COVID-19} 
\label{sec:inte_BC_iiot}

Blockchain-enabled IoMT can potentially solve the rising challenges amid COVID-19. Fig.~\ref{fig:scenario} presents an overview of solutions of blockchain-enabled IoMT to COVID-19. We next discuss these solutions in detail.

\subsection{Tracing pandemic origin}
It is crucial to trace the origin of pandemics like COVID-19 before the effective prevention and control countermeasures being made. Blockchain-enabled IoMT can offer an effective solution to this challenge, meanwhile offering a certain privacy protection of suspicious contacts. 

On the one hand, the wide adoption of thermal cameras and heat sensors in public places, such as hospitals, restaurants, and schools can offer early warnings. For example, it is reported in~\cite{udugama2020diagnosing} that thermal cameras have been adopted in airports and other transportation hubs to identify potentially-infected persons. The diverse thermal cameras and heat sensors are essentially connected with blockchain-enabled IoMT, which can guarantee data traceability and protect data privacy. Moreover, the introduction of thermal cameras to unmanned aerial vehicles (UAVs) can provide flexible surveillance on certain regions, which however may not be accessible to medical professionals~\cite{BChen:MCOM19}. On the other hand, the proliferation of nucleic acid tests and other COVID-19 test systems can help to identify infectious persons in a rapid manner. Thereafter, the interconnected IoMT systems can offer a ubiquitous data accessing services to various medical institutions, consequently identifying the origin of the virus.

\subsection{Quarantine and social distancing}
Both quarantine and keeping social distance are effective countermeasures to prevent a pandemic of contagious diseases. Blockchain-enabled IoMT can offer effective quarantine, isolation, and social distancing during outbreaks of contagious diseases such as COVID-19. 

In particular, many cities have also implemented compulsory quarantine for persons returning from overseas. However, the quarantine for massive inbound travelers may bring substantial burdens on public healthcare institutions during the outbreaks of the pandemics. Alternatively, home or hotel quarantine can be adopted, especially in densely populated cities. In this aspect, blockchain-based IoMT can offer a promising solution to the home or hotel quarantine, consequently reducing the loads on public medical resources. For example, a traveler who is required to wear a wristband since his/her arrival from overseas. The wristband can periodically send back the quarantined person's accurate location to disease control centers. Meanwhile, the introduction of blockchain and other cryptographic schemes can also protect the privacy of quarantined persons.

Moreover, the introduction of blockchain-based IoMT can also help to keep a social distance. For example, persons who wear specific wristband sensors (or tags) will be automatically alerted if their distance is too close (e.g., less than 6 feet). Furthermore, it is also reported that both Apple and Google released APIs for application developers to trace social distance and offer necessary warnings\footnote{\url{https://www.zdnet.com/article/the-worlds-first-contact-tracing-app-using-google-and-apples-api-goes-live/}}. The in-depth integration of these services with blockchain can further preserve user privacy and ensure data traceability.

\subsection{Smart hospital}
This COVID-19 crisis also reveals the potential vulnerabilities of existing medical institutions. In particular, hospitals are less resilient to the proliferation of infected persons. For example, many hospitals are incapable of handling normal pathologies since most doctors are summoned to treat COVID-19 patients. 

Blockchain-enabled IoMT can potentially address these issues. First, the introduction of diverse RFID tags and IoMT devices to hospitals can help to track and monitor the status of medical assets, such as the availability of beds and ambulances, breakdowns or malfunctions of medical devices (e.g., respirators). The status data can be stored in blockchain-enabled IoMT for further prescriptive and predictive analysis. Second, blockchain-enabled IoMT can also be deployed with hospital buildings to provide real-time monitoring on air quality, temperature, and environmental hygiene. Note that IoMT devices include not only passive devices (e.g., sensors and tags) but also active devices like robots, which are responsible for cleaning hospitals, disinfecting wards, or public places. Third, blockchain-enabled IoMT can also ensure the traceability and non-repudiation of IoMT data. For example, the sterilization of medical devices is crucial for the treatment of contagious diseases while the conventional sterilization records that are recorded by humans can be falsified. The introduction of blockchain-enabled IoMT can effectively address this issue due to the traceability and non-repudiation of blockchain data.

\subsection{Medical data provenance}

There are rising cases on fabrication or falsification of medical or healthcare data~\cite{herland2018big}. The fabricated data may affect the treatment of patients and cause a huge loss of medical insurance systems. Therefore, the \emph{data provenance} is introduced to improve the trustworthiness of medical data. The main initiative of data provenance is to store the meta-data of the medical data including the data source (origin), reproduction (derivation), and final clinic report covering the entire life cycle of medical data. 

Thanks to the traceability and immutability of blockchain, blockchain-enabled IoMT can well preserve \emph{data provenance}, consequently mitigating the fabrication risk of healthcare data~\cite{mackey2019fit}. In particular, the work~\cite{ramachandran2018smartprovenance} presents a blockchain-based data provenance system, which stores the encrypted data. Once there are any modifications to the stored data, the embedded smart contracts will be automatically triggered, consequently reporting the possible fabrications. Moreover, it is shown in~\cite{DLiu:JIOT20} that the blockchain-based framework is designed to improve the data provenance of IoT networks.

\subsection{Remote healthcare and telemedicine}
Both the aging population and limited medical resources bring burdens to conventional healthcare services. The outbreaks of pandemics make this situation more serious. Many senior citizens and chronic patients have to stay at home due to overwhelming patients in hospitals. 

Blockchain-based IoMT can potentially address this challenge. On the other hand, the proliferation of various IoMT wearable healthcare devices, such as embedded biosensors, medical sensors, and wristbands, leads to the feasibility of remote healthcare and telemedicine services at home. Thus, the advent of IoMT can potentially relieve the pressure of healthcare resources. For example, senior citizens in their homes or nursing centers can wear medical or healthcare devices, which can continuously take the measurements of their physiology information, such as heartbeat rate, blood pressure, and blood sugar. Through IoMT, doctors and their families can access their health records at any time and anywhere. In addition to the data generated from wearable devices, ambient information is also important because the environment factors (e.g., temperature and pressure) can also affect the analysis of doctors. Moreover, robots can also be leveraged for logistics (e.g., pill dispensers).

Despite the merits of IoMT, many healthcare and medical agencies refuse to adopt IoMT due to the risks of privacy breaches of patient data. In contrast to conventional IoMT, blockchain-based IoMT can well preserve the data privacy of patients and senior citizens via privacy-preservation schemes. Moreover, the appropriate authentication settings on top of blockchain can also eliminate the misuse (or abuse) of medical data~\cite{baza2020blockchain}. In this manner, unauthorized data access can be rejected. In summary, the introduction of blockchain-based IoMT can help to realize remote healthcare and telemedicine, consequently reducing the burdens at hospitals and saving the cost of scarce medical resources.

\textbf{Summary}. Table~\ref{tab:summary} summarizes the blockchain-enabled IoMT solutions to COVID-19. 

\begin{table*}[t]
\caption{Summary of blockchain-enabled IoMT solutions to COVID-19}
\label{tab:summary}
\renewcommand{\arraystretch}{1.75}
\centering
\begin{tabular}{|l|l|l|}
\hline
\multicolumn{1}{|c|}{\multirow{2}{*}{\bf Issues}} & \multicolumn{2}{c|}{\bf Blockchain-enabled IoMT solutions}  \\
\cline{2-3} 
& \multicolumn{1}{c|}{\bf IoMT}  & \multicolumn{1}{c|}{\bf Blockchain}  \\ \hline\hline
\textit{A. Tracing pandemic origin}                     & Thermal cameras, heat sensors and nucleic acid test & Traceability, privacy protection   \\\hline
\textit{B. Quarantine and social distancing}            & Wristband sensor and tags                           & Traceability, privacy protection   \\\hline
\textit{C. Smart hospital}                              & Medical asset management (RFID tags, sensors)       & Traceability, immutability         \\\hline
\textit{D. Medical data provenance}                     & IoT nodes, tags                                     & Traceability, anti-falsification   \\\hline
\textit{E. Remote healthcare and telemedicine}          & Wearable body sensors                               & Traceability, privacy preservation, authentication \\ \hline
\end{tabular}
\end{table*}


\section{Future directions}

Although blockchain-enabled IoMT is promising to combat the COVID-19 crisis, some challenges need to be addressed before the formal adoption of blockchain-enabled IoMT solutions in healthcare systems. We discuss several future directions in this section.

\subsection{Scalability of blockchain}
One of the main limitations of current blockchain systems is the poor scalability, which limits the wide adoption of blockchains in IoMT. The scalability of blockchain systems can be typically measured by transaction throughput, i.e., the number of transactions per second (tps). Compared with mature commercial payment systems, such as VISA and PayPal, which can reach 2,000 tps and 170 tps, respectively, Bitcoin (one of the most representative public blockchains) only has 7 tps. The throughput of public blockchains needs to be improved to handle the massive transactions and realtime requests in IoMT.

The countermeasures to the scalability of blockchain mainly include 1) leveraging more scalable consensus algorithms to improve the throughput; 2) adopting private or consortium blockchains, which are more scalable than public blockchains. With respect to 1), recent studies leverage the localization of the consensus strategies (such as sharding) to reach the consensus faster. Regarding the second types of solutions, private and consortium blockchains can process transactions faster than public blockchains mainly owing to the permissioned systems, which can reach the consensus more easily than public blockchains. Thus, private and consortium blockchains may be more suitable for IoMT. 

\subsection{Deep learning to enhance blockchain-enabled IoMT}
The massive available IoMT data contains huge values. Data analytics on the data generated by blockchain-enabled IoMT cannot only extract useful information but also enhance blockchain-enabled IoMT itself. We can identify potential system faults, vulnerabilities, and performance bottlenecks after analyzing IoMT data and blockchain data. However, conventional machine learning data analysis methods often require extensive feature engineering to extract features. The emerging deep learning algorithms that have less requirement on feature engineering can directly learn from the raw data. Therefore, deep learning approaches are more expected in the future for blockchain-enabled IoMT.

Meanwhile, the heterogeneity of IoMT data poses challenges on data analytics. Thus, different deep learning approaches may be adopted to handle different types of IoMT data. For example, deep convolutional neural networks (CNN) having the strengths in analyzing images can be used for biomedical images (such as X-Ray and MRI images) while recurrent neural networks (RNN) and the variants of RNN can be used for time-series data, such as electroencephalogram (EEG) data. In the future, the fusion of multiple deep learning algorithms will be an inevitable trend for data analytics in blockchain-enabled IoMT.

\subsection{Trustworthy AI for IoMT}
Despite the advances of deep learning and AI, they also bring trustworthy concerns, mainly on security and privacy vulnerabilities of AI models. In particular, it is reported in~\cite{Miller:Proc.IEEE2020} that AI models can be stolen, falsified, and poisoned (or contaminated). Meanwhile, IoMT data often needs to be outsourced to cloud servers which have the stronger computational capability to process and analyze IoMT data. However, cloud servers are often owned by untrusted third parties, which may accidentally or intentionally leak privacy-sensitive data to others. 

In this regard, the integration of blockchain and AI might be a solution to this issue. In particular, blockchain systems may effectively authenticate and authorize data accessing when IoMT data is outsourced to cloud servers. Meanwhile, the authentication and authorization mechanisms can be also applied to the management of cloud servers which are required to be validated through the decentralized blockchain system. Moreover, the recent advances in federated learning algorithms bring opportunities to solve this issue from a different perspective. Instead of uploading IoMT data to remote cloud servers, federated learning algorithms~\cite{Fadlullah:TETC20} can train AI models from the collected IoMT data locally at edge nodes, which are deployed at base stations or IoT gateways in approximation to users. Thereafter, the trained AI models can be merged together. In this manner, the risk of privacy leakage is reduced since no IoMT data is transferred to cloud servers which are often owned by untrusted third parties. It will be a future direction to integrate federated learning with blockchain, thereby further enhancing the trust of IoMT systems.




    
    

\section{Conclusion}

We have been experiencing an unprecedented crisis caused by COVID-19 while the recent ICT advances also bring opportunities to combat this crisis. Specifically, the wide adoption of the Internet of medical things (IoMT) in healthcare institutions can help to collect massive medical and healthcare data, which can be used by medical practitioners to diagnose and identify diseases, consequently offering appropriate treatments. Therefore, the introduction of IoMT can undertake the burdens of healthcare systems while IoMT also is also faced with security and privacy vulnerabilities. On the other hand, blockchain technologies can potentially enhance security and protect the privacy of IoMT systems. 

Therefore, this article presents a study on the integration of blockchain and IoMT amid the COVID-19 crisis. In this article, we first give an overview of blockchain and IoMT. We then put forth a multi-layer framework for blockchain-enabled IoMT. Moreover, we discuss the solutions brought by blockchain-enabled IoMT to COVID-19 from five perspectives such as tracing the pandemic origin, quarantining and social distancing, smart hospital, medical data provenance, and remote healthcare and telemedicine. We finally outline the future directions in blockchain-enabled IoMT. We believe that we can finally win the battle against COVID-19 with the aid of blockchain-enabled IoMT and other technologies. 

\ifCLASSOPTIONcaptionsoff
  \newpage
\fi

\bibliographystyle{IEEEtran}
\bibliography{ref}

\begin{thebibliography}{10}
\providecommand{\url}[1]{#1}
\csname url@samestyle\endcsname
\providecommand{\newblock}{\relax}
\providecommand{\bibinfo}[2]{#2}
\providecommand{\BIBentrySTDinterwordspacing}{\spaceskip=0pt\relax}
\providecommand{\BIBentryALTinterwordstretchfactor}{4}
\providecommand{\BIBentryALTinterwordspacing}{\spaceskip=\fontdimen2\font plus
\BIBentryALTinterwordstretchfactor\fontdimen3\font minus
  \fontdimen4\font\relax}
\providecommand{\BIBforeignlanguage}[2]{{%
\expandafter\ifx\csname l@#1\endcsname\relax
\typeout{** WARNING: IEEEtran.bst: No hyphenation pattern has been}%
\typeout{** loaded for the language `#1'. Using the pattern for}%
\typeout{** the default language instead.}%
\else
\language=\csname l@#1\endcsname
\fi
#2}}
\providecommand{\BIBdecl}{\relax}
\BIBdecl

\bibitem{MShen:MNET19}
M.~{Shen}, Y.~{Deng}, L.~{Zhu}, X.~{Du}, and N.~{Guizani},
  ``{Privacy-Preserving Image Retrieval for Medical IoT Systems: A
  Blockchain-Based Approach},'' \emph{IEEE Network}, vol.~33, no.~5, pp.
  27--33, 2019.

\bibitem{dai-iotj-2019}
H.-N. {Dai}, Z.~{Zheng}, and Y.~{Zhang}, ``Blockchain for internet of things: A
  survey,'' \emph{IEEE Internet of Things Journal}, vol.~6, no.~5, pp.
  8076--8094, 2019.

\bibitem{ZZheng:FGCS2020}
Z.~Zheng, S.~Xie, H.-N. Dai, W.~Chen, X.~Chen, J.~Weng, and M.~Imran, ``{An
  Overview on Smart Contracts: Challenges, Advances and Platforms},''
  \emph{Future Generation Computer Systems}, vol. 105, pp. 475--491, April
  2020.

\bibitem{Ahmed:MC20}
M.~{Ahmed} and A.~K. {Pathan}, ``{Blockchain: Can It Be Trusted?}''
  \emph{Computer}, vol.~53, no.~4, pp. 31--35, 2020.

\bibitem{Qadri:CST20}
Y.~A. {Qadri}, A.~{Nauman}, Y.~B. {Zikria}, A.~V. {Vasilakos}, and S.~W. {Kim},
  ``The future of healthcare internet of things: A survey of emerging
  technologies,'' \emph{IEEE Communications Surveys \& Tutorials}, vol.~22,
  no.~2, pp. 1121--1167, 2020.

\bibitem{Boudagdigue:TIFS20}
C.~{Boudagdigue}, A.~{Benslimane}, A.~{Kobbane}, and J.~{Liu}, ``Trust
  management in industrial internet of things,'' \emph{IEEE Transactions on
  Information Forensics and Security}, pp. 1--1, 2020.

\bibitem{udugama2020diagnosing}
B.~Udugama, P.~Kadhiresan, H.~N. Kozlowski, A.~Malekjahani, M.~Osborne, V.~Y.
  Li, H.~Chen, S.~Mubareka, J.~B. Gubbay, and W.~C. Chan, ``{Diagnosing
  COVID-19: the disease and tools for detection},'' \emph{ACS nano}, vol.~14,
  no.~4, pp. 3822--3835, 2020.

\bibitem{BChen:MCOM19}
B.~{Chen}, M.~{Imran}, N.~{Nasser}, and M.~{Shoaib}, ``{Self-Aware Autonomous
  City: From Sensing to Planning},'' \emph{IEEE Communications Magazine},
  vol.~57, no.~4, pp. 33--39, 2019.

\bibitem{herland2018big}
M.~Herland, T.~M. Khoshgoftaar, and R.~A. Bauder, ``Big data fraud detection
  using multiple medicare data sources,'' \emph{Journal of Big Data}, vol.~5,
  no.~1, p.~29, 2018.

\bibitem{mackey2019fit}
T.~K. Mackey, T.-T. Kuo, B.~Gummadi, K.~A. Clauson, G.~Church, D.~Grishin,
  K.~Obbad, R.~Barkovich, and M.~Palombini,
  ``‘fit-for-purpose?’--challenges and opportunities for applications of
  blockchain technology in the future of healthcare,'' \emph{BMC medicine},
  vol.~17, no.~1, pp. 1--17, 2019.

\bibitem{ramachandran2018smartprovenance}
A.~Ramachandran and M.~Kantarcioglu, ``{SmartProvenance: a distributed,
  blockchain based data provenance system},'' in \emph{Proceedings of the
  Eighth ACM Conference on Data and Application Security and Privacy}, 2018,
  pp. 35--42.

\bibitem{DLiu:JIOT20}
D.~{Liu}, J.~{Ni}, C.~{Huang}, X.~{Lin}, and X.~{Shen}, ``{Secure and Efficient
  Distributed Network Provenance for IoT: A Blockchain-based Approach},''
  \emph{IEEE Internet of Things Journal}, pp. 1--1, 2020.

\bibitem{baza2020blockchain}
M.~Baza, M.~M. Fouda, M.~Nabil, A.~T. Eldien, H.~Mansour, and M.~Mahmoud,
  ``Blockchain-based distributed key management approach tailored for smart
  grid,'' in \emph{Combating Security Challenges in the Age of Big Data}.\hskip
  1em plus 0.5em minus 0.4em\relax Springer, 2020, pp. 237--263.

\bibitem{Miller:Proc.IEEE2020}
D.~J. {Miller}, Z.~{Xiang}, and G.~{Kesidis}, ``{Adversarial Learning Targeting
  Deep Neural Network Classification: A Comprehensive Review of Defenses
  Against Attacks},'' \emph{Proceedings of the IEEE}, vol. 108, no.~3, pp.
  402--433, 2020.

\bibitem{Fadlullah:TETC20}
Z.~M. {Fadlullah} and N.~{Kato}, ``{HCP: Heterogeneous Computing Platform for
  Federated Learning Based Collaborative Content Caching Towards 6G
  Networks},'' \emph{IEEE Transactions on Emerging Topics in Computing}, pp.
  1--1, 2020.

\end{thebibliography}

\begin{IEEEbiographynophoto}{Hong-Ning Dai} is currently with Faculty of Information Technology at Macau University of Science and Technology as an associate professor. He obtained the Ph.D. degree in Computer Science and Engineering from Department of Computer Science and Engineering at the Chinese University of Hong Kong. His current research interests include Internet of Things and blockchain technology. He has served as editors for IEEE Access, Ad Hoc Networks and Connection Science, guest editors for IEEE Transactions on Industrial Informatics, IEEE Transactions on Emerging Topics in Computing and IEEE Open Journal of the Computer Society. 
\end{IEEEbiographynophoto}
	
\begin{IEEEbiographynophoto}{Muhammad Imran} is working as an associate professor in the College of Applied Computer Science, King Saud University. His research interests include mobile and wireless networks, Internet of Things, cloud and edge computing, and information security. He has published more than 200 research articles in reputable international conferences and journals. His research is supported by several grants. He serves as an associate editor for many top ranked international journals. He has received various awards.
\end{IEEEbiographynophoto}

\begin{IEEEbiographynophoto}{Noman Haider} is a working as a lecturer in the College of Engineering and Science at Victoria University, Australia. He received his Ph.D. in engineering from the University of Technology Sydney, Australia in 2019. His research interests include mobile and wireless networks, Internet of Things, data science, and information security.
\end{IEEEbiographynophoto}

\end{document}